\documentclass[twocolumn,prb,superscriptaddress,showpacs]{revtex4}

\usepackage[dvips]{graphicx}

\newcommand\nc[2]{\newcommand#1{#2}}
\nc{\idx[1]}{{\rm #1}}
\nc{\be}{\begin{equation}}
\nc{\ee}{\end{equation}}
\nc{\bea}{\begin{eqnarray}}
\nc{\eea}{\end{eqnarray}}
\nc{\ds}{\displaystyle}
\def\ff#1#2{\raisebox{0.13ex}{{\footnotesize ${\displaystyle\f{#1}{#2}}$}}}
\nc{\qre}{r_\idx{e}}
\nc{\qrp}{r_\idx{p}}
\nc{\bR}{\mbox{{\boldmath $R$}}}
\nc{\br}{\mathbf{r}}
\nc{\pout}{p_\idx{out}}
\nc{\holerad}{R_0}
\nc{\aB}{a_\idx{B}}
\nc{\qme}{m_\idx{e}}
\nc{\qAA}{\, \mbox{\AA}}
\nc{\qAa}{\, \mbox{\scriptsize{\AA}}}
\nc{\qH}{{\cal H}}
\nc{\qd}{\hbox{d}}
\nc{\p}{\partial}
\def\f#1#2{{#1\over#2}}
\nc{\eV}{\:\mbox{eV}}
\nc{\aJ}{\,\mbox{aJ}}

\def\eg{{\it e.g.,\ }}
\def\ie{{\it i.e.,\ }}

\def\etal{{\it et al.}}

\def\cf{cf.\ }

\nc{\RhaB}{\frac{\holerad}{\aB}}
\nc{\RhaBB}{\left(\frac{\holerad}{\aB}\right)}
\nc{\RhaBBB}{\left(\ff{\holerad}{\aB}\right)}
\def\free{\idx{free-PS}}
\def\qEfree{E_\free}
\def\qPsifree{\Psi_\free}
\def\qEnum{E^\idx{(num)}}
\def\qEunit{|\qEfree|}
\def\szazalek{\%\,}
\def\const{\hbox{\sl const.\ }}
\begin{document}

\title{On the inner structure of confined positronium}
\author{Tam\'as F\"ul\"op}
\affiliation{Institute of Particle and Nuclear Studies,
  High Energy Accelerator Research Organization (KEK),
  Tsukuba, Ibaraki 305-0801, Japan}
\author{Z\'en\'o Farkas}
\affiliation{Institute of Physics, University Duisburg-Essen,
  D-47048 Duisburg, Germany}
\author{Alfred Seeger}
\author{J\'anos Major}
\affiliation{Max-Planck-Institut f\"ur Metallforschung,
  Heisenbergstr.\ 3, D-70569 Stuttgart, Germany}

\date{April 16, 2003}

\begin{abstract}
  We study the inner structure of a positronium confined to a void
  by treating it as a two-particle quantum system where the electron
  is captured by an infinite potential well and the positron feels
  only the Coulomb attraction of the electron. We calculate the ground
  state and the related quantities that govern the annihilation rate,
  and discuss implications of the results.
\end{abstract}


\maketitle

\section{Introduction}

Shortly after the discovery of the positron (e$^+$) \cite{anderson33},
it was recognized that e$^+$ may form an `atom' (e$^+$e$^-$) which,
from the point of view of spectroscopy, may be looked upon as an
ultralight isotope of hydrogen \cite{mohorovicic34}.
With the same right, (e$^+$e$^-$) may be considered as an ultralight 
isotope of antihydrogen. Together with the photon, the Majorana neutrino,
and the electrically neutral $\pi$-meson ($\pi ^0$), (e$^+$e$^-$) 
thus belongs to those `particles' that are their own `antiparticles'. 
In 1945 (e$^+$e$^-$), at that time still hypothetical, was named
positronium \cite{ruark45}, later to be equipped with the `chemical' 
symbol Ps~ \cite{mcgervey59}. At about the same time its
main properties were established \cite{wheeler46}. Among these are that Ps
is rather short-lived due to the particle--antiparticle annihilation and 
that its ground state splits into two substates, a singlet state 
l$^1$S$_0$ and a triplet state l$^3$S$_1$.
         
In the l$^1$S$_0$ state, Ps is called {\it parapositronium} (p-Ps), and
in the l$^3$S$_1$ state, {\it orthopositronium} (o-Ps).
Under most circumstances, the energy excess of o-Ps over p-Ps, 
$\Delta E_{\rm o-p} = 0.8412 \times 10^{-4}$~eV, is negligibly small, 
hence the ratio of the production rates of o-Ps and p-Ps
is usually equal to that of the statistical weights of the substates,
namely 3:1. The most important differences between o-Ps and p-Ps result 
from the different annihilation modes of these two substates, which, 
in turn, are consequences of the conservation of energy and angular
momentum. The {\it self-annihilation} of p-Ps results in the generation of 
{\it two} $\gamma$-quanta of equal energies 
$E_{\gamma} = m_{\rm e} c^2 = 0.511$~MeV, 
where $m_{\rm e}$ is the electron mass and $c$ the speed of light in
vacuum, whereas the self-annihilation of o-Ps requires the generation 
of, at least, {\it three} $\gamma$-quanta. Owing to the smallness of the
coupling constant of quantum electrodynamics ($\alpha \approx
1/137$) and to various
statistical factors, the ratio between the 2$\gamma$ annihilation rate, 
$\Gamma_{2 \gamma} = 7.99 \times 10^9$~s$^{-1}$, 
and the 3$\gamma$ annihilation rate, 
$\Gamma_{3 \gamma} = 7.04 \times 10^6$~s$^{-1}$, 
exceeds $10^{3}$ and is therefore highly significant.

The preceding statements refer to Ps in vacuum or in extremely dilute
gases. In dense gases and in condensed matter the fate of o-Ps is
radically changed, since now the e$^+$--e$^-$ annihilation may 
(and in general does) involve electrons other than the 1s 
electrons of the Ps atoms. This leads to 2$\gamma$ annihilation by 
the `pick-off' mechanism. This mechanism may be visualized as follows. 
An electron in the neighbourhood with spin direction
opposite to that of the e$^+$ `picks off' the positron from a Ps atom
and annihilates it by a 2$\gamma$ reaction. The pick-off
reaction rate $\Gamma_{\rm po}$ is determined by the probability 
density of the electrons with opposite spin at the e$^+$ location. 
This density is clearly less than that of electrons in the 1s orbit of
Ps, in condensed matter typically by about a factor of ten. Hence 
$\Gamma_{\rm po}$ is smaller than the self-annihilation rate 
$\Gamma_{2\gamma}$ but still large compared to $\Gamma_{3\gamma}$. 
This has several important consequences:
\begin{list}{}{\setlength{\topsep}{0pt}\setlength{\itemsep}{0pt}
\setlength{\parsep}{0pt}}      
\item[(i)] Positron annihilation in condensed matter takes 
place virtually exclusively by the generation of two $\gamma$ quanta.
\item[(ii)]  The annihilation rate of o-Ps depends 
significantly on the environment in which the positronium atom is located.
\item[(iii)] Because of (ii), 
the o-Ps annihilation rate may be used
as indicator of the preferred sites of Ps in condensed matter.
\item[(iv)] In Ps-forming solids containing imperfections 
that attract Ps, the site-sensitivity of $\Gamma_{\rm po}$
may be used to detect and monitor such imperfections.  
\end{list} 

The pick-off annihilation rates of o-Ps fall into a range that allows
accurate measurements of their inverses, the o-Ps lifetimes 
\cite{hautojarvi79}. Often it is possible to perform e$^+$
{\it lifetime spectrometry}, i.e., to distinguish different positron
annihilation 
sites by the different e$^+$ lifetimes associated with them. 
In order to gain a theoretical understanding of the site
dependence of the e$^+$ lifetimes, one has to calculate the
wavefunctions of Ps atoms that are confined to the various imperfections
that are capable of trapping Ps. This has turned out to be
a very formidable task.

The electrostatic forces experienced by electrons and positrons are
equal and opposite. In most situations the variation of the
electrostatic potentials over the diameter of a Ps atom
(about two Bohr radii, see below) is negligibly small compared with the 
positronium binding energy $\qEfree$
(in vacuum about 6.8~eV). Unless a Ps atoms enters a kind of hydrogen 
bond, the electrostatic forces on its constituents cancel to an
excellent approximation. Then the interaction of Ps with its 
condensed-matter environment is entirely due to the exchange
interaction between the Ps-electrons and the host electrons since 
for e$^+$ implanted into matter
the Pauli principle is irrelevant. The fact that the ensuing repulsion
of Ps by the host atoms
is short-ranged suggests that the confinement of Ps to interstices 
or locations where
atoms are missing may be represented by a rectangular potential well at
these sites.

Since even this model is still a nontrivial quantum mechanical problem,
Tao and Eldrup initiated a further simplified model in which the
positronium is considered as a point particle in a spherical infinite
potential well, being in the ground state \cite{tao72,chuang73,eldrup81}, 
extending the ideas of Ferrell
\cite{ferrell57} and Brandt \etal\ \cite{brandt60}. In this
model, pick-off annihilation happens in a layer $ r_0 \le r \le \holerad
$ near the surface of the spherical volume of radius $\holerad$,
the layer representing the medium outside the hole. The width of
this layer is a parameter fitted experimentally. Later, this model
was refined by considering a potential wall of finite height
\cite{nakanishi88} and of unsharp shape \cite{mukherjee98}.
In these modifications, the pick-off annihilation of the positron
occurs in the region $ \holerad \le r < \infty $, which seems to be
a better approximation from the physical point of view and provides
a better fit to the experimental data. Nevertheless, as stressed
by Mukherjee \etal\ \cite{mukherjee98},
this approach is still an oversimplification
since it treats the positronium as a single point particle.

In this paper, we provide an approach that avoids this simplification.
We determine
the quantum mechanical ground state of the positron-electron
two-particle system with an infinite potential well that acts only
on the electron, in addition to the Coulomb attraction between the
positron and the electron. Hence, we give account of the arising
nontrivial internal structure of a confined positronium.

Due to the complicated nature of the problem at hand,
we calculate the ground state numerically, in terms of an expansion
with respect to an appropriate set of basis functions (truncated
to contain the first 216 basis functions) and via a variational
method, for various pore sizes. The probability of finding the positron
outside the void and the overlap of the positron and the electron inside
the hole are also determined. We analyse and interpret the results, and
discuss how further refinements of the model and the explanation of the
observed temperature dependence of the pick-off rate can be implemented.

It is to be noted that a related investigation was done by
Sommerfeld and Welker \cite{sommerfeld38},
where the ground-state energy and wave
function of a {\it hydrogen atom} in a hole is calculated. In that
study, the proton is considered to be infinitely heavy and to reside
in the center of
the spherical hole, and the electron is confined to the cavity by
an infinite potential well. We make a comparison between those
results and our ones, and provide a simple common explanation
of the behavior of both systems.

\section{The model}

Using the nonrelativistic framework and observing that, under the
present conditions, spatial and spin degrees of freedom are decoupled, 
we formulate our model expounded above with a normalized electron--positron
two-particle scalar wave function $\Psi ({\br_{\rm e}}, {\br_{\rm p}})$,
the Hamiltonian
    \be
\qH = - \frac{\hbar^2}{2\qme}
(\nabla_{\br_{\rm e}}^2 + \nabla_{\br_{\rm p}}^2) -
\frac{e^2}{4\pi\varepsilon_0}
\frac{1}{\left|{\br_{\rm e}} - {\br_{\rm p}}\right|} \, ,
    \ee
and the boundary condition
    \be
\Psi ({\br_{\rm e}}, {\br_{\rm p}})
|_{\mbox{\raisebox{-1.8pt}{\scriptsize $\qre = \holerad$}}} = 0 \, ,
\qquad \forall {\br}_{\rm p} \, .
    \label{eq:bc} \ee
Here, $\br_{\rm e}$ denotes the electron co-ordinates and $\br_{\rm p}$
the positron co-ordinates, $R_0$ is the radius of the
potential well to which the electron is confined, 
$e$ is the elementary charge, $\; \varepsilon_0 =
8.854187 \cdot 10^{-12}$~As/Vm \hskip .7ex is the dielectric constant
of the vacuum, and $ 2 \pi \hbar = h $ is Planck's constant.

We wish to determine the ground-state wave function --- from now on,
$\Psi ({\br_{\rm e}}, {\br_{\rm p}})$ will refer to the ground
state --- together with the ground-state energy, and to calculate the
following two ground-state related quantities.
First, the probability
    \be
\pout = \int\limits_{r_p\ge\holerad} \hspace{-1.5ex} \qd^3 {\br_{\rm p}}
\hspace{-0.5ex} \int\limits_{\qre \le \holerad} \hspace{-1.5ex} \qd^3
{\br_{\rm e}} \, |\Psi({\br_{\rm e}}, {\br_{\rm p}})|^2
    \label{eq:pout} \ee
that the positron will be found outside the hole. This quantity is a global
measure of the pick-off annihilation rate of o-Ps.
Second, the electron--positron contact parameter
    \be
\kappa = \frac{\int\limits_{\qre \le \holerad} \hspace{-1.5ex} \qd^3
{\br_{\rm e}} |\Psi({\br_{\rm e}}, {\br_{\rm p}} = {\br_{\rm
e}})|^2 \hspace{0.5ex} } {\int \qd^3 {\br_{\rm e}} |\qPsifree
({\br_{\rm e}}, {\br_{\rm p}} = {\br_{\rm e}})|^2} \, ,
    \label{eq:kappa} \ee
where $\qPsifree$ is the ground-state wave function of the free
positronium, having the form
    \be
\frac{1}{\sqrt{8 \pi \aB^3}} \;\; {\rm exp}
\left(-\frac{\left|{\br}_{\rm p}-{\br}_{\rm e}\right|}{2 \aB}\right)
    \label{eq:free-ps} \ee
after separating the centre-of-mass motion,
with $\aB = 4 \pi \epsilon_0 \hbar^2/\left(m_{\rm e}e^2\right) =
0.529 \times 10^{-10}$~m denoting the Bohr radius.
This quantity measures the overlap of the elecron and the positron,
and is equal to the ratio of the self-annihilation rate of confined
Ps to that of free Ps.

Because of the spherical symmetry of the system, the ground-state
wave function depends on three independent scalar variables only.
For these we choose the distance of the electron from the hole
centre, $\qre = |{\br}_{\rm e}|$, the electron--positron separation
$R = |{\br}_{\rm p}-{\br}_{\rm e}|$, and the angle $\chi$
between ${\br}_{\rm e}$ and ${\br}_{\rm p}-{\br}_{\rm e}$,
satisfying $\cos \chi = {\br}_{\rm e} \cdot ({\br}_{\rm p}-{\br}_{\rm e})
/ (\qre R)$.

A lengthy yet straightforward calculation shows that, in terms of
these variables, the Hamiltonian reads
\begin{widetext}
  \bea
  \qH & = & - \frac{\hbar^2}{2m_{\rm e}} \left(
  \f{\p^2}{\p \qre^2} + 2 \, \f{\p^2}{\p R^2} +
  \frac{ 2 \qre^2 + R^2 + 2 \qre R \cos \chi }{ \qre^2 R^2 } \,
  \f{\p^2}{\p \chi^2}
  - 2 \cos \chi \, \f{\p^2}{\p \qre \, \p R} + 2 \, \frac{\sin \chi}{R}
  \, \f{\p^2}{\p \qre \, \p \chi}   \right. \nonumber \\
  & & \left. 
  + 2 \, \frac{\sin \chi}{\qre}
  \f{\p^2}{\p R \, \p \chi}
  + \f{2}{\qre} \, \f{\p}{\p \qre} + \frac{4}{R} \,
  \f{\p}{\p R} + \frac{ 2 \qre^2 \cos \chi + R^2 \cos \chi +
    2 \qre R }{ \qre^2 R^2 \sin \chi } \, \f{\p}{\p \chi} \right)
  - \frac{e^2}{4\pi\varepsilon_0} \frac{1}{R}
  \label{eq:hamilton}
  \eea
\end{widetext}
The Hamiltonian (\ref{eq:hamilton}) is not separable and does not
contain a small parameter on which a perturbation expansion might
be based, therefore, we perform the subsequent calculations
numerically.

\section{The method of computation}

We use a variational method to compute the approximate ground-state
wave function, minimizing the energy $\langle \Psi, \qH \Psi \rangle$
in the function space
    \be
\Psi = \sum\limits_{n=1}^N C_n \Psi_n \, ,
    \label{eq:phisum} \ee
where the functions $\Psi_n (n = 1, \dots, N)$ form a set of
appropriately chosen orthonormal base functions. The minimization
is carried out by finding the eigenvector of the matrix of the
elements $\qH_{mn} = \langle \Psi_m, \qH \Psi_n \rangle$ with the
lowest eigenvalue. The components of this eigenvector are then
identified with the coefficients $C_n$ of the approximate ground
state wave function (\ref{eq:phisum}). The base functions are
chosen in such a way that
\\ (i)
the $\qre$-dependence of the wave function is similar to the ground
state of a {\it single electron} in an infinitely deep potential well,
    \be
\psi_{\rm e} (\qre) = \const \f{ \sin (\pi \qre / \holerad) }{\qre}
    \label{eq:freepsie} \ee
(see, \eg the work of Galindo and Pascual \cite{galindo90}),
\\ (ii)
the $R$-dependence resembles that of the free positronium
(\cf Eq.~\ref{eq:free-ps}), and
\\ (iii)
that the matrix elements can be computed relatively fast.\\
To this end, we introduce the functions
   \begin{eqnarray}
\varphi^{(1)}_i (\qre) = & (\qre^2-\holerad^2)^i \, \; \;
                                       &  (i = 1, \, \ldots, \, N_1), \\
\varphi^{(2)}_j (R)    = & R^{j-1} {\rm e}^{-R/\xi}
                                       &  (j = 1, \, \ldots, \, N_2),
\label{eq:atoms} \\
\varphi^{(3)}_k (\chi) = & \cos^{k-1}\chi
                                       &  (k = 1, \, \ldots, \, N_3),
    \end{eqnarray}
from which, by means of the Gram--Schmidt orthonormalization procedure
\cite{schmidt07}, we obtain the orthonormalized functions $\, \psi^{(1)}_i
(\qre) \,$, $\, \psi^{(2)}_j (R) \,$ and $\, \psi^{(3)}_k (\chi) \,$,
respectively. The products
    \be
\Psi_{ijk} (\qre, R, \chi) = \psi^{(1)}_i (\qre) \cdot
\psi^{(2)}_j (R) \cdot \psi^{(3)}_k (\chi)
    \label{eq:Psiprod} \ee
serve as the base functions $\Psi_n$,
{\em i.e.}, the expansion (\ref{eq:phisum}) is implemented as
    \be
\Psi = \sum\limits_{i=1}^{N_1} \sum\limits_{j=1}^{N_2}
\sum\limits_{k=1}^{N_3} C_{ijk} \Psi_{ijk} \, .
    \label{eq:Psisum} \ee

The variational method is improved by considering $\xi$ as an
additional, nonlinear adjustable parameter, with respect to which
the energy is also minimized (for a free positronium, $ \xi = 2
\aB  = 1.058 \times 10^{-10}$~m).

For the first functions $\, \psi^{(1)}_1 (\qre) \,$, $\, \psi^{(2)}_1
(R) \,$ and $\, \psi^{(3)}_1 (\chi) \,$, the orthonormalization procedure
means only a normalization. It is plausible to expect that the product
    \be
\Psi_{111} = \const \cdot \varphi^{(1)}_1 (\qre) \cdot
\varphi^{(2)}_1 (R) \cdot \varphi^{(3)}_1 (\chi)
\label{eq:product}
   \ee
proves to be the dominant basis function in the ground state
(\ref{eq:Psisum}).

The matrix components $\qH_{mn}(\holerad,\xi)$ were calculated
analytically by the software Maple V~ \cite{maple}
and subsequently 
exported to the format of programming language C, and the numerical
solution of the eigenproblem, as a function of $\holerad$ and $\xi$,
was performed by a C program using routines from Numerical Recipes
\cite{press92}. For a fixed $\holerad$, the lowest eigenvalue was
minimized as a function of $\xi$, and the eigenvector corresponding
to this eigenvalue was determined. The lowest eigenvalue gave
the approximate energy of the ground state, while the components of
the eigenvector provided the coefficients $C_{ijk}$ of the approximate
ground-state function (\cf Eq.~\ref{eq:Psisum}). Finally, using
the resulting $C_{ijk}$ values, $\pout$ and $\kappa$ were calculated
again by Maple V. The computations took approximately two months on
a workstation.

\section{Results}

It was found that a base-function set with $ N_1 = N_2 = N_3 = 2
$ sufficed for determining the energy, $ N_1 = N_2 = N_3 = 3 $ for
$\pout$, and $ N_1 = N_2 = N_3 = 4$ for $\kappa$ --- in the sense
that any further increase of the number of base functions produced
only a negligible change in the values of $E$, $\pout$, or $\kappa$.
The results presented below were calculated using the base-function
set $ N_1 = N_2 = N_3 = 6 $. The calculations were performed for
33 different values of the hole size in the range $ \aB \le \holerad
\le 98 \aB$.


\begin{figure}
\begin{center}
\includegraphics[angle=-90, width=0.95\linewidth]{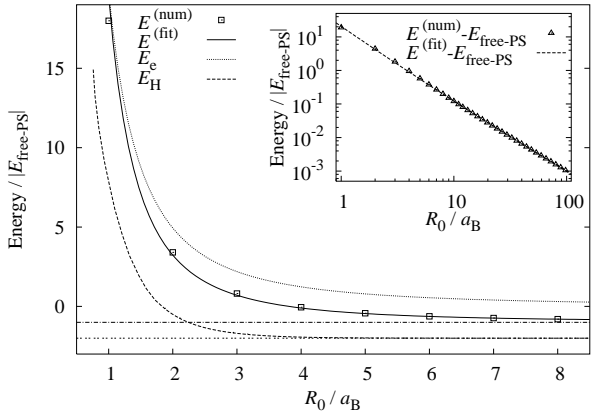}
\caption{The calculated ground-state energy of the confined positronium,
$\qEnum$, as the function of the hole radius.
The data points are fitted by the
function given in Eq.\ (\ref{eq:energy_fit}) (solid line). The dotted
line shows the ground-state energy of a single electron confined to a hole,
$ E_\idx{e} = \pi^2 \hbar^2 / (2 \qme \holerad^2)$,
 and the dashed line denotes the ground-state energy of a hydrogen atom
in a hole. The ground-state energy of the free positronium, $\qEfree$,
and the ground-state energy of the free hydrogen atom, $2\qEfree$, are
also indicated. Inset: the difference between the
ground-state energies of the confined and free positronium on a log-log
scale, together with the fit (\ref{eq:energy_fit}).
}
\label{fig:energy1}
\end{center}
\end{figure}

The resulting ground-state energy $E$ is shown in Fig.\ \ref{fig:energy1}
as the function of the hole radius. For large $\holerad$ values, $E$
approaches the ground-state energy $\qEfree = -6.80 \eV \,$
of the free positronium. The
dependence of the energy on $\holerad$ is well described by the function
    \be
E = \qEunit \left[ 8.75 \RhaBBB^{-2} + 11.78 \RhaBBB^{-2.53} - 1 \right].
\label{eq:energy_fit}
    \ee
The error of the numerical parameters in this fitted function
is less then 3.7\szazalek.  For comparison,
Fig.\ \ref{fig:energy1} displays the ground-state energy of a single
{\it electron} confined to a hole, $ E_\idx{e} = \pi^2 \hbar^2 /
(2 \qme \holerad^2) $ (see, \eg the work of Galindo and Pascual 
\cite{galindo90}), 
and the
ground-state energy of a {\it hydrogen atom} in a hole with its
proton in the centre (reproduced from the work of Sommerfeld and Welker 
\cite{sommerfeld38}) as well.

\begin{figure}
\begin{center}
\includegraphics[angle=-90, width=0.95\linewidth]{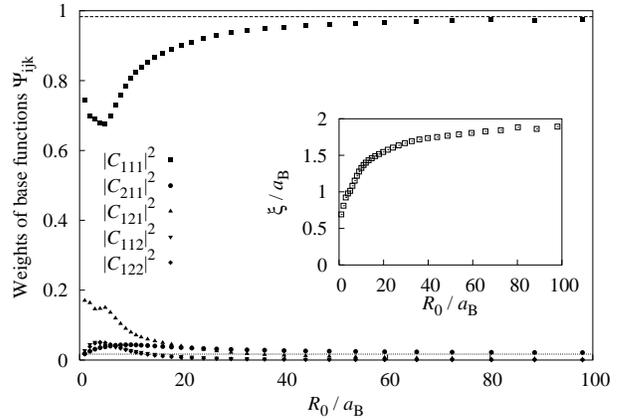}
\caption{The weights of the most dominant base functions in the ground
state, as functions of the hole radius. (Concerning the coefficients
themselves, $C_{111}$ and $C_{121}$ are positive, and $C_{211}$,
$C_{112}$ and $C_{122}$ are negative, for all $\holerad$.) The weights of
$\psi^{(1)}_1 (\qre)$ and $\psi^{(1)}_2 (\qre)$ in the ground state of
a single electron in a hole (\ref{eq:freepsie}) --- approximately $0.983$
and $0.017$, respectively --- are also displayed. Inset: the parameter
$\xi$, as the function of the hole radius. The value of $\xi$
for the free positronium is $2 \aB$.
}
\label{fig:Cxi}
\end{center}
\end{figure}

The determined ground state $\Psi$ of the positronium in a pore is
characterized by the coefficients $C_{ijk}$ and $\xi$, Fig.\ \ref{fig:Cxi}
presents the most dominant coefficients and $\xi$, as the function of
the hole radius. The coefficients are displayed in their squared form,
$|C_{ijk}|^2$, which give the physical weights of the corresponding base
functions $\Psi_{ijk}$ in the ground state $\Psi$.
All coefficients $C_{ijk}$ were found real, as was to be expected.


\begin{figure}
\begin{center}
\includegraphics[angle=-90, width=0.95\linewidth]{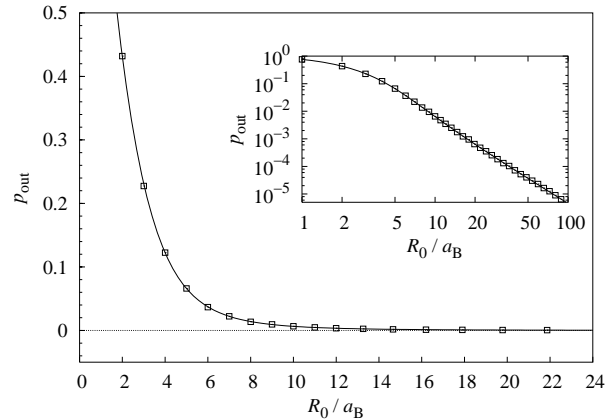}
\caption{The probability $\pout$ that the positron of the confined
positronium can be found outside the hole, as the function of the
hole radius. The data points are fitted by the function given in Eq.\
(\ref{eq:pout_fit}).
}
\label{fig:pout}
\end{center}
\end{figure}

The calculated $\holerad$-dependence of the probability $\pout$
that the positron can be found outside the void is shown in
Fig.\ \ref{fig:pout}. The data may be fitted by the function
    \be
\pout = \frac{1 + 0.018 \RhaBB^2}{1 + 0.34 \RhaBB^2 + 0.004 \RhaBB^5}
    \label{eq:pout_fit} \ee
with an uncertainty of the coefficients of less than 4\szazalek.


\begin{figure}
\begin{center}
\includegraphics[angle=-90, width=0.95\linewidth]{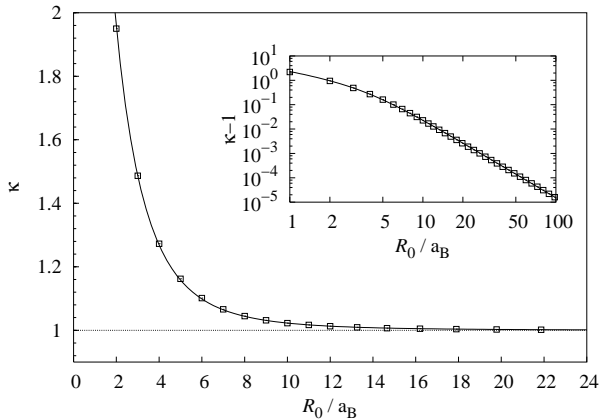}
\caption{The density parameter $\kappa$ of the confined positronium
as the function of the hole radius. The data points are fitted by
the function given in Eq.\ (\ref{eq:kappa_fit}).
}
\label{fig:kappa}
\end{center}
\end{figure}

Similarly, the results for the density parameter $\kappa$ (\cf Fig.\
\ref{fig:kappa}) may be described by the function
    \be
\kappa = 1 + \frac{1}{0.41 \RhaB + 0.024 \RhaBB^{3.22}}
    \label{eq:kappa_fit} \ee
with uncertaities of the fit parameters of less then 2\szazalek.


\section{Analysis of the results}

The first observation to make is that, as expected, for large hole radii,
the system tends to the free positronium. This can be seen on each the
quantities $E$, $\xi$, $\pout$ and $\kappa$. Also, as expected, it is
$\Psi_{111}$ among the base functions $\Psi_{ijk}$ whose presence is
the strongest in the determined wave function, and this dominance
increases with the size of the hole to almost $100\szazalek$.

\begin{figure}
\begin{center}
\includegraphics[angle=-90, width=0.95\linewidth]{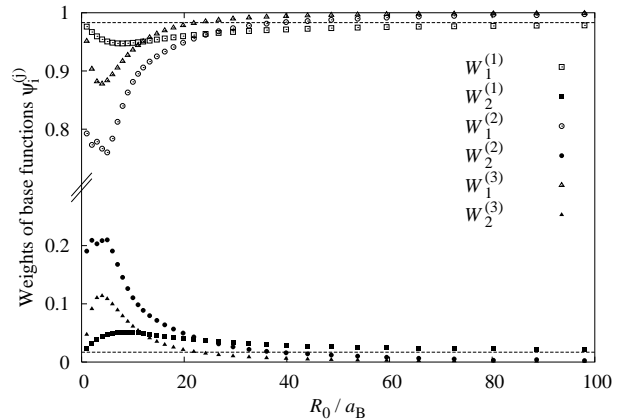}
\caption{The weights of the base functions $\psi^{(1)}_1 (\qre)$,
$\psi^{(1)}_2 (\qre)$, $\psi^{(2)}_1 (R)$, $\psi^{(2)}_2 (R)$,
and $\psi^{(3)}_1 (\chi)$, $\psi^{(3)}_2 (\chi)$
in the ground state of a confined positronium.
For each of the three variables, the presence of the
higher order base functions is small, beyond the fourth
order they can be practically neglected even for small hole sizes.
The dashed lines show the weights of functions
$\psi^{(1)}_1 (\qre)$ and $\psi^{(1)}_2 (\qre)$
in the ground state of a single confined electron
(\cf Fig.~\ref{fig:Cxi}).
}
\label{fig:weights}
\end{center}
\end{figure}

To investigate the ground state more closely, let us study its dependence
on the variables $\qre$, $R$ and $\chi$. This can be done by calculating
the weights of the base function components $ \psi^{(1)}_i (\qre) $, $
\psi^{(2)}_j (R) $ and $ \psi^{(3)}_k (\chi) $ in the wave function. These
weights are defined as
    \be
W^{(1)}_i = \sum\limits_{j=1}^{N_2} \sum\limits_{k=1}^{N_3} |C_{ijk}|^2
    \label{eq:w} \ee
for a $ \psi^{(1)}_i (\qre) $, and analogously for a $ \psi^{(2)}_j (R)
$ and a $ \psi^{(3)}_k (\chi) $. The calculated weights are shown
on Fig.~\ref{fig:weights}, as functions of the hole radius.


Concerning the $\qre$-dependence, we can observe that the weights
$W^{(1)}_i$ are almost independent of $\holerad$, and that they are very
close to the weights of $ \psi^{(1)}_i (\qre) $s in the
ground state of a single electron in a hole [\cf Eq.~(\ref{eq:freepsie})],
     \be
|c_1|^2 = \f{945}{\pi^3} \approx 0.983 \, , \hskip 0.5ex
|c_2|^2 = \f{93555 (10 - \pi^2)^2}{\pi^{10}} \approx 0.017
\label{eq:c1c2} 
     \ee
(the higher weights $|c_3|^2,\ |c_4|^2,\ \ldots$ are negligible so it is
enough to consider the first two ones). Therefore, the $\qre$-dependent
part of $\Psi$ is close to the wave function of the confined electron,
even for small values of $\holerad$, and with a better and better accuracy
as $\holerad$ increases.

The $R$-dependence is more influenced by $\holerad$. In a small void,
the components higher than the first one contribute up to $20\szazalek$
and this amount decreases to zero as the hole size is increased. For
large $\holerad$, only the first component remains, and in the meantime
its parameter $\xi$ approaches the value of the free positronium.
The $\chi$-dependence contains a $10\szazalek$ amount of the higher base
functions for small $\holerad$, which decreases to zero for increasing
hole radius. Since $\, \psi^{(3)}_1 (\chi) = \const $, this means that,
for larger pore sizes, the ground state becomes $\chi$-independent.

Turning to the ground-state energy, a simple interpretation of its
found $\holerad$-depend\-ence can be that the presence of the finite
hole increases the energy of the electron by the amount of $ E_\idx{e}
= \pi^2 \hbar^2 / (2 \qme \holerad^2) $, similarly as it does with a
single electron. Fig.\ \ref{fig:energy2} shows that this interpretation is
fairly good. In Appendix A, we show that, with an only somewhat more
involved physical argument, it is possible to give an even more precise
approximation to the energy as the function of $\holerad$.


\begin{figure}
\begin{center}
\includegraphics[angle=-90, width=0.95\linewidth]{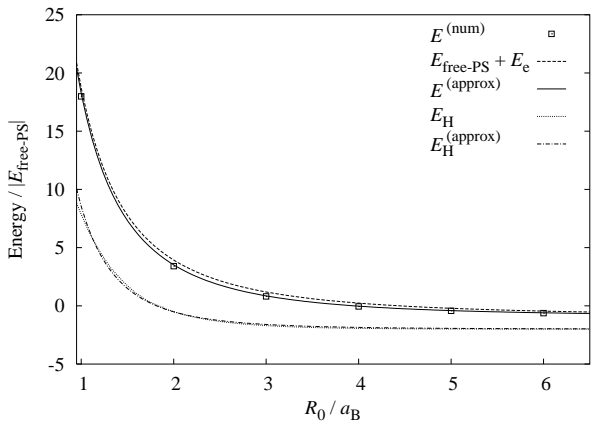}
\caption{The ground-state energy of the confined positronium as
a function of the hole radius, together with two approximations
to it, $ \qEfree + \pi^2 \hbar^2 / (2 \qme \holerad^2) $ and
Eq.~\ref{eq:app9}. The ground-state energy of the confined hydrogen
atom as a function of the hole radius, together with its approximation
Eq.~\ref{eq:app11}, is also shown.
}
\label{fig:energy2}
\end{center}
\end{figure}

The probability that the positron can be found outside the hole, $\pout$,
is one for holes much smaller than $\aB$, since the characteristic space
region needed for the positron is of the order of $\aB$, like in the free
positronium (see Appendix A as well on this subject). On the other side,
for large hole sizes, $\pout$ falls down to zero as $\holerad^{-3}$
(\cf Eq.~\ref{eq:pout_fit}). This latter can also be understood, an
explanation for this is presented in Appendix B.

The density parameter $\kappa$ also behaves as expected. For $\holerad$
large, it is 1 since the positronium is almost free. By gradually
squeezing the hole, the allowed space for the electron becomes smaller,
it can no more go so far away from the positron as in a free positronium.
This results in some increasing in the overlap-measuring $\kappa\,$. We
can guess what happens when $\holerad$ becomes much smaller than the
Bohr radius: The electron is practically confined at the origin, the
wave function of the positron becomes similar to what it is like in
the antihydrogen atom, and, therefore, the $R$-dependent part of $\Psi$
becomes
    \be
\f{1}{\sqrt{\pi \aB^3}} \;\; {\rm exp} \left(-\f{R}{\aB}\right) \, .
    \label{eq:anti} \ee
The $\kappa$ value corresponding to this wave function is 8 [\cf
Eqs.~(\ref{eq:kappa}) and (\ref{eq:free-ps})] so the increase that can be
observed in Fig.~\ref{fig:kappa} is expected to slow down for $\holerad <
\aB$ and to converge to the value 8 reached at $\holerad = 0$.

We mention that a similar behavior of $\kappa$ is expected if one
replaces the infinite potential well by some other confining
potential. Limiting the electron to a region is plausible to cause
a kind of squeezing the two particles onto each other, irrespective
of the concrete form of the confining mechanism.

\section{Discussion}

In the framework of our model, the lifetime of an orthopositronium is
determined by the pick-off annihilation rate, proportional to
$\pout$, and the intrinsic $3\gamma$ decay rate of the orthopositronium,
which is considered to be its vacuum value multiplied by $\kappa$.
The angular correlation of the two emitted photons of the pick-off
process can be obtained by generalizing the Fourier approach of
Mukherjee and coworkers \cite{mukherjee98}
to the total momentum of the present two-particle system.

The infinite potential well, chosen in our calculations presented
here, can be a good approximation for the confining mechanism for
the electron in the pore for some materials and less good for
others. The same variational method can be applied for other
potential shapes, modifying only the electronic base functions, in
the appropriate way. Naturally, finding and using the corresponding
electronic base functions for various potential shapes may make
the calculations hard to carry out in practice. In this respect,
it is worth mentioning a feasible alternative approach for this
purpose. It is known from quantum mechanics
that the boundary condition of requiring the vanishing
of the wave function $\Psi$, Eq.~\ref{eq:bc}, is not the only possibility
for confining a particle to a finite region \cite{fulop00}.
There exists actually a
one-parameter family of quantum mechanically allowed boundary
conditions, of the form $\ \Psi + L \, \p \Psi / \p \qre = 0 \ $,
that is, the vanishing of a combination of the wave function and
its derivative, where $L$ is an arbitrary (real) length scale
parameter. The well-known and most frequently used case, the
vanishing of the wave function itself is simply the special case $L = 0$.
Instead of choosing different potential shapes it is simpler to choose
only a different value of $L$ in the boundary condition. The
corresponding electronic base functions can still be chosen as
polynomials, fulfilling now the new boundary condition. One
can expect that the precise shape of the confining potential is not so
important in practice. Bearing in mind that these boundary
conditions express nothing but that the quantum probability
cannot flow outside from the hole\cite{fulop00},
and that any confining potential shape wishes to ensure the
same --- although not completely strictly ---, we can guess that all
confining potentials will, in their effect, resemble one or another boundary
condition (potential well) with an appropriate $L$. Therefore, the
possibilities provided by the general potential shapes may be fairly well
represented with the one-parameter family of sharp potential wells.

In this paper, we have calculated the ground state only. To give
account of the observed temperature dependence of the pick-off
decay rate, it would be useful to determine
the excited states, too. Indeed, as demonstrated by Goworek and coworkers
\cite{goworek98}
and Gidley and coworkers \cite{gidley99} 
for some variants of the Tao-Eldrup model,
this temperature dependence can be attributed to the fact that it
is also possible for the positronium to be in one of its excited states,
approximately with the thermal equilibrial probability $e^{-E/kT}$.
Explaining the effect in our framework, the excited states possess
different $\pout$ and, consequently, a different annihilation rate.
Calculating the excited states is also possible in the variational
method, after suitable modifications.

In the end we mention that it would be interesting to perform a
calculation similar to the one presented here for systems when the
two particles are of different mass, \eg when the positron is
replaced with a proton or a $\mu^+$ particle, as an interpolation between
the positron and the infinitely heavy limit discussed by Sommerfeld and
Welker. Unfortunately, our
results cannot be simply 'renormalized' to be applicable to such
situations. The reason for this is that, for different masses, the
Hamiltonian has a more general form than Eq.~(\ref{eq:hamilton}) has,
and these differences are not only some easily rescalable numerical
factors but mean a more general and complicated structure of the
Hamiltonian. However, the considerations of Section 4 and the
Appendices may more easily be adjusted to the case of different
masses, and in this way at least some approximate information could
be obtained for those systems.

\begin{acknowledgments}
We wish to thank Nikolay Djourelov for useful references.
\end{acknowledgments}

\appendix

\section{Estimating the ground state energy}

A closer, yet not involved interpretation of the found
$\holerad$-dependence of the energy of the positronium in a hole is
motivated by the simple derivation of the ground-state energy of the free
hydrogen atom. There, based on the uncertainty relation, one approximates
the average momentum of the electron by $p = \hbar / r$, where $r$
is the size of the average space region ``run'' by the electron. Thus the
energy of the electron is
    \be
\f{p^2}{2\qme} - \frac{e^2}{4\pi\varepsilon_0} \f{1}{r} = \f{\hbar^2}
{2\qme} \f{1}{r^2} - \frac{e^2}{4\pi\varepsilon_0} \f{1}{r} \, .
    \label{eq:app1} \ee
Minimizing this expression in $r$ one reaches just the correct
ground-state energy, and the corresponding $r$ is also nothing else than
the Bohr radius $\aB$.

As a next step, let us consider the free positronium. There, the
average distance between the electron and the positron is $2\aB$ (\cf
Eq.~\ref{eq:free-ps}). Neglecting the motion of the center-of-mass,
the average region run by the effective particle of reduced mass $\qme/2$
belonging to the relative motion is then $2\aB$. The energy is, therefore,
considered as
    \be
\f{\hbar^2}{2(\qme/2)} \f{1}{(2\aB)^2} - \frac{e^2}{4\pi\varepsilon_0}
\f{1}{2\aB} \, ,
    \label{eq:app2} \ee
which proves to be just the correct energy value $\qEfree$. If we return
to the 'two particles'-picture, the kinetic energy in Eq.~(\ref{eq:app2}) is
shared by the two particles so each possesses the half of it,
    \be
\f{1}{2} \; \f{\hbar^2}{2(\qme/2)} \f{1}{(2\aB)^2} = \f{\hbar^2}{2\qme}
\f{1}{(2\aB)^2} \, .
    \label{eq:app3} \ee
Using this in the reverse way, we find that the characteristic space
region for both particles is $2\aB$.

Now, when we confine the electron in a hole, its characteristic region
will be influenced by the hole as well. To this end, let us determine
the characteristic space size of a single electron in a hole with the same
logic as before, from its known energy:
    \be
E_\idx{e} = \f{\pi^2 \hbar^2}{2 \qme \holerad^2} = \f{\hbar^2}{2\qme}
\f{1}{(\holerad/\pi)^2}
    \label{eq:app4} \ee
``implies'' that the corresponding characteristic distance is
$\holerad/\pi$. \cite{footnoteA}

When the electron is both confined to a hole and attracted by a positron,
its allowed space region will be determined by the smaller of the two
corresponding length scales --- except for some narrow intermediate
region when the two length scales are equal or similar. A simple as
well as reasonable formula for estimating such a joint length scale
from two ones is
    \be
\f{1}{l_\idx{joint}^2} = \f{1}{l_1^2} + \f{1}{l_2^2} \, ,
    \label{eq:app5} \ee
which in our case reads
    \be
\f{1}{\qre^2} = \f{1}{(2\aB)^2} + \f{1}{(\holerad/\pi)^2} \, .
    \label{eq:app6} \ee
Concerning the space region of the positron, we make the simplest choice
to consider it unaltered with respect to the case of the free positronium,
\ie to be $2\aB$. The average distance between the electron and the
positron, $r_\idx{ep}$, also has to be estimated: For this we can use
the formula
    \be
r = \sqrt{r_1^2 + r_2^2}
    \label{eq:app7} \ee
as being a ``half-way'' between the minimal distance $|r_1-r_2|$ and the
maximal one $r_1+r_2$~:
    \bea
|r_1 - r_2| & = & \sqrt{r_1^2 - 2 r_1 r_2 + r_2^2} \le \sqrt{r_1^2 + r_2^2}
\nonumber \\
& \le & \sqrt{r_1^2 + 2 r_1 r_2 + r_2^2} = r_1 + r_2
    \label{eq:app8} \eea
($r_1, \: r_2 \ge 0$). Putting all these together, we estimate the energy as
    \bea
E^\idx{(approx)} & = & K_\idx{e} + K_\idx{p} + V(r_\idx{ep}) =
\f{\hbar^2}{2\qme} \f{1}{\qre^2} + \f{\hbar^2}{2\qme} \f{1}{(2\aB)^2}
\nonumber \\
& & - \f{e^2}{4\pi\varepsilon_0} \f{1}{\sqrt{ (\qre)^2 + (2\aB)^2 }} \, .
    \label{eq:app9} \eea
As can be seen on Fig.\ \ref{fig:energy2}, this formula provides a pretty
good approximation for the ground-state energy of the confined positronium.

We can apply the same style of approach for the case of the hydrogen atom
in a hole as well. There, the joint length scale of the length scale $\aB$
of the free hydrogen and the one of the hole, $\holerad/\pi$, is given by
    \be
\f{1}{r_\idx{H}^2} = \f{1}{\aB^2} + \f{1}{(\holerad/\pi)^2} \, .
    \label{eq:app10} \ee
In the potential energy this same length scale will appear, since the
proton stands in the center of the hole. Consequently, our estimate for
the energy of a hydrogen atom in a hole is
    \be
E_\idx{H}^\idx{(approx)} = \f{\hbar^2}{2\qme} \f{1}{r_\idx{H}^2} -
\f{e^2}{4\pi\varepsilon_0} \f{1}{r_\idx{H}} \, .
    \label{eq:app11} \ee
Fig.\ \ref{fig:energy2} shows that our reasoning gives a fairly good
approximation for this system as well.

\section{The probability $\pout$ for large holes}

An explanation for the found large-$\holerad$ asymptotic behavior
$\pout \sim \holerad^{-3}$ can be given by the following argument:
Let us rewrite Eq.~(\ref{eq:pout}) as
    \be
\pout = 8 \pi^2 \int\limits_{0}^{\infty} R^2 \, \qd R \;
\int\limits_{0}^{\holerad} \qre^2 \: \qd \qre \int\limits_{\chi_1}^{\chi_2}
\sin \! \chi \,
\qd \chi \; |\Psi(\qre, R, \chi)|^2 \, ,
    \label{eq:integral} \ee
where the integration boundaries $\chi_1 \ge 0$ and $\chi_2 \le \pi$
for $\chi$ are determined from the condition $ \qrp = \sqrt{ \qre^2 +
R^2 + 2 \qre R \cos \! \chi }\, \ge \holerad $.
Expanding $|\Psi(\qre, R, \chi)|^2$ corresponding to
Eq.~(\ref{eq:Psisum}) and also expanding all $\psi$s in it in terms of
$\varphi$s, let us examine the integral of one term,
    \be
\varphi^{(1)}_i (\qre) \, \varphi^{(1)}_{i'} (\qre) \cdot
\varphi^{(2)}_j (R) \, \varphi^{(2)}_{j'} (R) \cdot \varphi^{(3)}_k
(\chi) \, \varphi^{(3)}_{k'} (\chi) \, .
    \label{eq:szorzat} \ee
Concerning the variable $\qre$, the main contribution to the integral
must come from $\,\qre \approx \holerad\,$, within the range $\,
|\qre - \holerad| \sim \xi \,$, since otherwise $R$ has to be around
$\holerad$ to ensure $\, \qrp \ge \holerad \,$, but that is exponentially
suppressed by $ e^{-2R/\xi} \approx e^{-2\holerad/\xi} $ in the
integrand. Now, for $\qre \approx \holerad$ and $R \ll \holerad$,
the condition $\,\qrp \ge \holerad\,$ is fulfilled by $\chi$ in the
region $\, 0 \le \chi \le \pi/2 \,$. The integral of $\: \sin \! \chi \:
\varphi^{(3)}_{k} (\chi) \, \varphi^{(3)}_{k'} (\chi) \,$ over this
region gives an $\holerad$-independent numerical factor. Turning to
the $\qre\,$-integration, for $\,\qre \approx \holerad\,$, the factor
$\: \qre^2 \: \varphi^{(1)}_i (\qre)\: \varphi^{(1)}_{i'} (\qre) \:$
behaves as
    \bea
\qre^2 \, (\qre^2 - \holerad^2)^{i+i'} & = & \qre^2 \, (\qre +
\holerad)^{i+i'} (\qre - \holerad)^{i+i'} 
\nonumber \\
& \approx & \holerad^2 \: (2
\holerad)^{i+i'} (\qre - \holerad)^{i+i'}
    \label{eq:asre} \eea
so its integral within the range $\, 0 \le |\qre - \holerad| \sim \xi \,$ is
proportional to $\holerad^{i+i'+2} \xi^{i+i'+1}$. The normalizing factors
standing before $ \varphi^{(1)}_i (\qre) $ and $ \varphi^{(1)}_{i'}
(\qre) $ bring in a factor $\holerad^{- (2i + 2i' + 3)}$, resulting
in a behavior $\holerad^{- (i + i' + 1)}$. The last integration,
with respect to $R$, is an independent one,
    \be
\int\limits_{0}^{\infty} R^2 \, \qd R \;
\varphi^{(2)}_j (R) \, \varphi^{(2)}_{j'} (R) \, ,
    \label{eq:Rint} \ee
with a result depending only on $\xi$  so it does not modify the
$\holerad$-asymptotics (note that, for large $\holerad$, $\xi$ is
independent of the hole size, it tends to the constant $2\aB$). Hence,
we can see that the strongest $\holerad$-asymptotics among the terms
in $\pout$ is caused by the $i = i' = 1$ term, and is found to be
$\holerad^{-3}$.


\begin{thebibliography}{22}
\bibitem[{foo()}]{footnoteA}
\bibinfo{note}{It is reasonable that this distance is smaller than $\holerad$
  since, in a hole with an impenetrable wall around it, the particle is not
  allowed to extend beyond $\holerad$ while, for a hydrogen atom, the electron
  can go further than the Bohr radius.}
\end{thebibliography}
\end{document}